\documentclass[12pt]{article}

\RequirePackage{cite}
\RequirePackage{times}
\RequirePackage{fullpage}
\RequirePackage{ifthen}
\RequirePackage{lscape}

\usepackage{soul}

\makeatletter
\def\@cite#1#2{$^{\mbox{\scriptsize #1\if@tempswa , #2\fi}}$}
\makeatother

\newcommand{\spacing}[1]{\renewcommand{\baselinestretch}{#1}\large\normalsize}
\spacing{2}

\def\@maketitle{%
  \newpage\spacing{1}\setlength{\parskip}{12pt}%
    {\Large\bfseries\noindent\sloppy \textsf{\@title} \par}%
    {\noindent\sloppy \@author}%
}

\newenvironment{affiliations}{%
    \setcounter{enumi}{1}%
    \setlength{\parindent}{0in}%
    \slshape\sloppy%
    \begin{list}{\upshape$^{\arabic{enumi}}$}{%
        \usecounter{enumi}%
        \setlength{\leftmargin}{0in}%
        \setlength{\topsep}{0in}%
        \setlength{\labelsep}{0in}%
        \setlength{\labelwidth}{0in}%
        \setlength{\listparindent}{0in}%
        \setlength{\itemsep}{0ex}%
        \setlength{\parsep}{0in}%
        }
    }{\end{list}\par\vspace{12pt}}

\renewenvironment{abstract}{%
    \setlength{\parindent}{0in}%
    \setlength{\parskip}{0in}%
    \bfseries%
    }{\par\vspace{-4pt}}

\newenvironment{addendum}{%
    \setlength{\parindent}{0in}%
    \small%
    \begin{list}{Acknowledgements}{%
        \setlength{\leftmargin}{0in}%
        \setlength{\listparindent}{0in}%
        \setlength{\labelsep}{0em}%
        \setlength{\labelwidth}{0in}%
        \setlength{\itemsep}{12pt}%
        }
    }
    {\end{list}\normalsize}

\usepackage{times}
\usepackage{subfigure}

\usepackage{epstopdf}
\usepackage{graphicx}
\usepackage{epsfig}

\def \aap{Astron.~Astrophys.}
\def \apjl{Astrophys.~J.}
\def \apj{Astrophys.~J.}

\def \mnras{Mon.~Not.~Roy.~Astron.~Soc.}

\def \nat{Nature}
\def\msun{{\,M_\odot}}

\newcommand{\lsim}{\,\rlap{\raise 0.35ex\hbox{$<$}}{\lower 0.7ex\hbox{$\sim$}}\,}
\newcommand{\gsim}{\,\rlap{\raise 0.35ex\hbox{$>$}}{\lower 0.7ex\hbox{$\sim$}}\,}

\begin{document}

\noindent
 {\large {\bf {\fontfamily{phv}\selectfont Giant Magnetized Outflows from the Centre of the Milky Way
}}}

\noindent
Ettore Carretti,$^{1}$ Roland M. Crocker,$^{2,3}$
Lister Staveley-Smith$^{4,5}$, Marijke Haverkorn$^{6,7}$,
Cormac Purcell$^{8}$,
B.~M. Gaensler$^{8}$,
Gianni Bernardi$^{9}$,
Michael~J. Kesteven$^{10}$, and
Sergio Poppi$^{11}$

\begin{affiliations}
 \item CSIRO Astronomy and Space Science, PO Box 276, Parkes, NSW 2870, Australia
 \item Max-Planck-Institut f{\" u}r Kernphsik, P.O. Box 103980 Heidelberg, Germany
 \item Research School of Astronomy \& Astrophysics, Australian National University, Weston Creek, ACT 2611, Australia
 \item International Centre for Radio Astronomy Research, M468, University of Western Australia, Crawley, WA 6009, Australia
 \item ARC Centre of Excellence for All-Sky Astrophysics
 \item Department of Astrophysics/IMAPP, Radboud University Nijmegen, P.O. Box 9010, 6500 GL Nijmegen, The Netherlands
 \item Leiden Observatory, Leiden University, P.O. Box 9513, 2300 RA Leiden, The Netherlands
 \item Sydney Institute for Astronomy, School of Physics, The University of Sydney, NSW 2006, Australia
 \item Harvard--Smithsonian Center for Astrophysics, 60 Garden Street, Cambridge, MA, 02138, USA
 \item CSIRO Astronomy and Space Science, PO Box 76, Epping, NSW 1710, Australia
 \item INAF � Osservatorio Astronomico di Cagliari, St. 54 Loc. Poggio dei Pini, I-09012 Capoterra (CA), Italy
\end{affiliations}

\begin{abstract}
The nucleus of the Milky Way  is known to harbour regions of intense star formation activity as well as a super-massive black hole\cite{Morris1996}. 
Recent  Fermi space telescope observations have 
revealed 
regions of $\gamma$-ray emission 
reaching far above and 
below the Galactic Centre, the so-called Fermi bubbles\cite{Su2010}. 
It is uncertain whether these were generated by 
nuclear star formation or by quasar-like
outbursts of the central black hole\cite{Su2012,Zubovas2011,Crocker2011a,Crocker2011c}
and no information on the structures'
magnetic field has been reported.
Here we report on the detection of two giant, 
linearly-polarized radio Lobes, containing three ridge-like
 sub-structures, emanating from the Galactic Centre. 
The Lobes each extend  $\sim 60^\circ$, 
bear a close correspondence to the Fermi bubbles,  are located in the Galactic bulge, and
are permeated by strong magnetic 
 fields of up to 15 $\mu$G.
Our data 
signal that the radio Lobes  originate in a bi-conical, star-formation 
(rather than black hole) 
driven 
outflow from the Galaxy's central 200 pc that transports a massive magnetic energy of 
 $\sim 10^{55}$ erg into the Galactic halo.
The ridges 
wind around this outflow and, we suggest,  constitute a `phonographic' record of nuclear star formation activity over at least 10 Myr.
\end{abstract}

We use the images of the recently concluded S-band Polarization All Sky Survey (S-PASS) that has mapped the polarized radio emission of the entire southern sky with the Parkes Radio Telescope at a frequency of 2307~MHz, with 184~MHz bandwidth, and 9' angular resolution\cite{Carretti2011}. 

\begin{landscape}
\begin{figure}
\centering
  \includegraphics[angle=0, width=0.9\hsize]{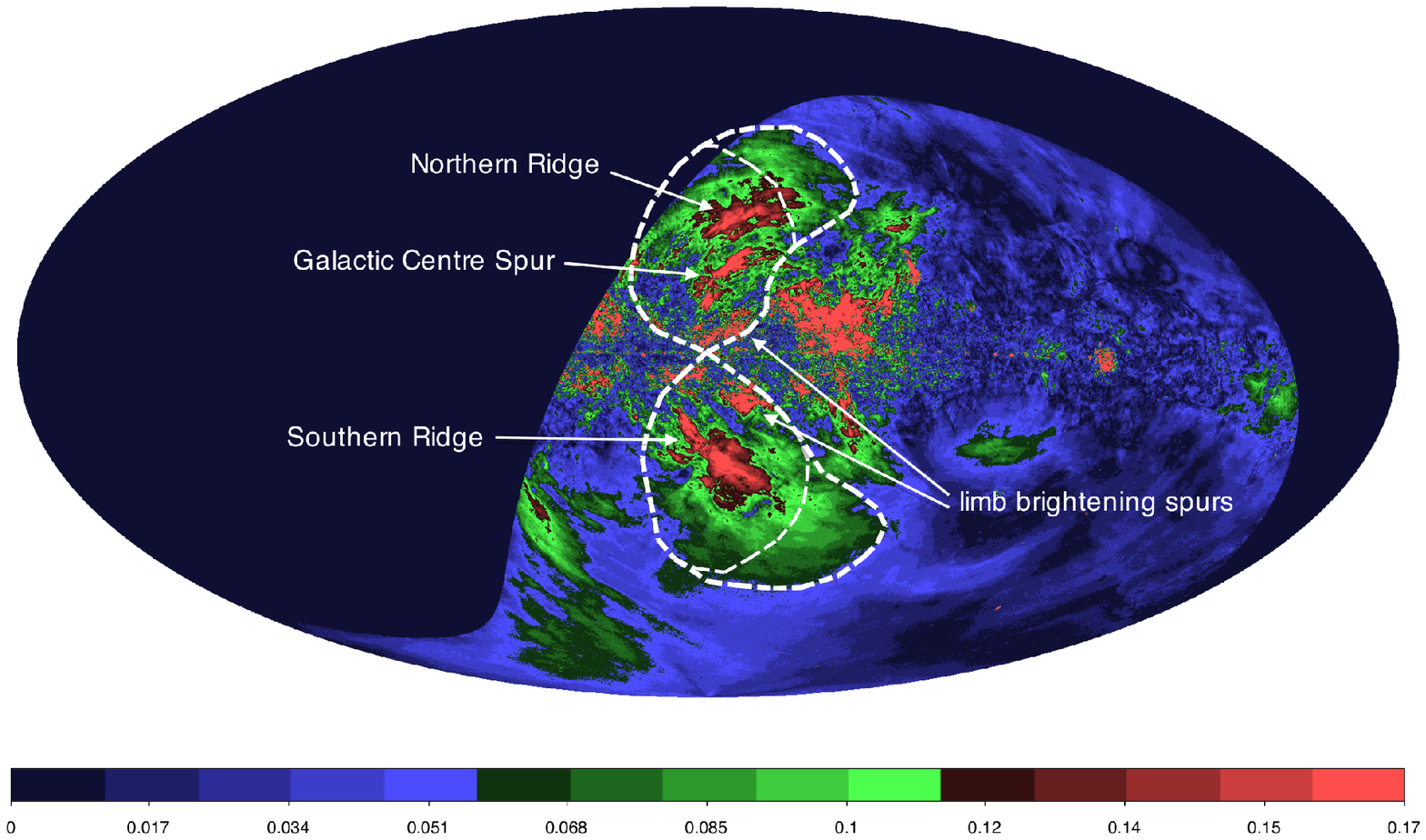}
\caption{    
 \label{fig_1}
 }
\end{figure}
\end{landscape}

\begin{figure}
\centering
    \includegraphics[angle=0, width=0.9\hsize]{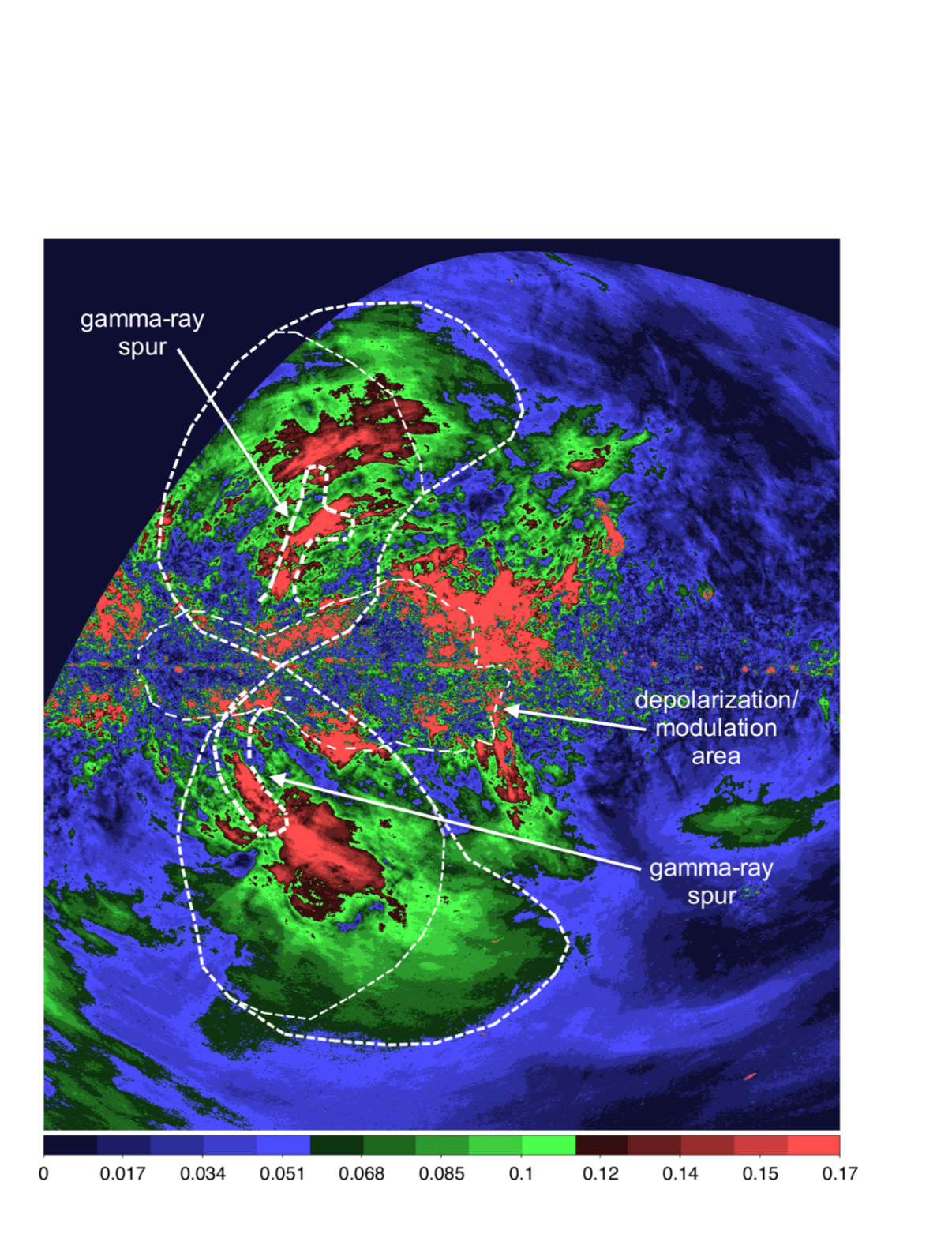}
\caption{   
\label{lobes:Fig}
}
\end{figure}

\begin{landscape}
\begin{figure}
\centering
  \includegraphics[angle=0, width=1.0\hsize]{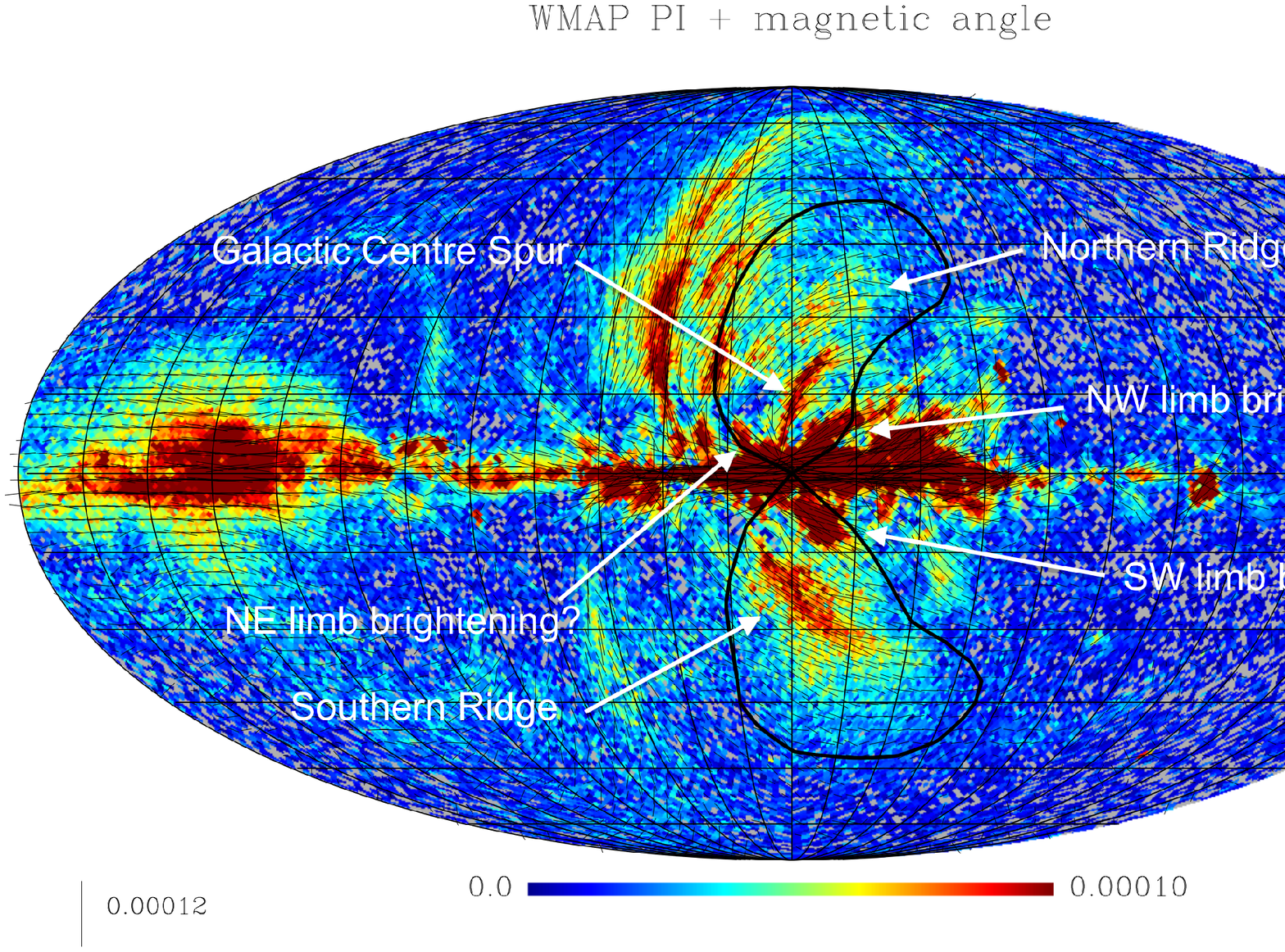}
\caption{
 \label{wmap:Fig}
}
\end{figure}
\end{landscape}

The Lobes   exhibit diffuse polarized emission (Figure~\ref{fig_1}), an integrated total intensity  flux of 21~kJy, 
and high polarization fraction of 25\%.  
They trace the Fermi Bubbles excepting  the top western (i.e., right) corners where they  extend beyond the $\gamma$-ray region. 
Depolarization by HII
regions establish that the Lobes are almost certainly associated with the Galactic Centre  (Figure~\ref{lobes:Fig} and Supplementary Information), implying their height is $\sim$8 kpc.
Archival data of WMAP[\cite{Hinshaw2009}] reveal the same structures at the microwave frequency of 23~GHz (Figure~\ref{wmap:Fig}). 
The 2.3 to 23~GHz spectral index $\alpha$ (with flux density $S$ modelled as $S_\nu \propto \nu^{\alpha}$) of linearly-polarized emission interior to the Lobes spans the range $\alpha$~=~[-1.0, -1.2]
 generally steepening with projected distance from the Galactic plane (see the Supplementary Information). 
Along with the high polarization fraction, this phenomenology 
indicates the Lobes are due to cosmic ray electrons, 
transported from the plane,  synchrotron-radiating 
in a partly ordered  magnetic field. 

\begin{figure}
\centering
  \includegraphics[angle=0, width=1.0\hsize]{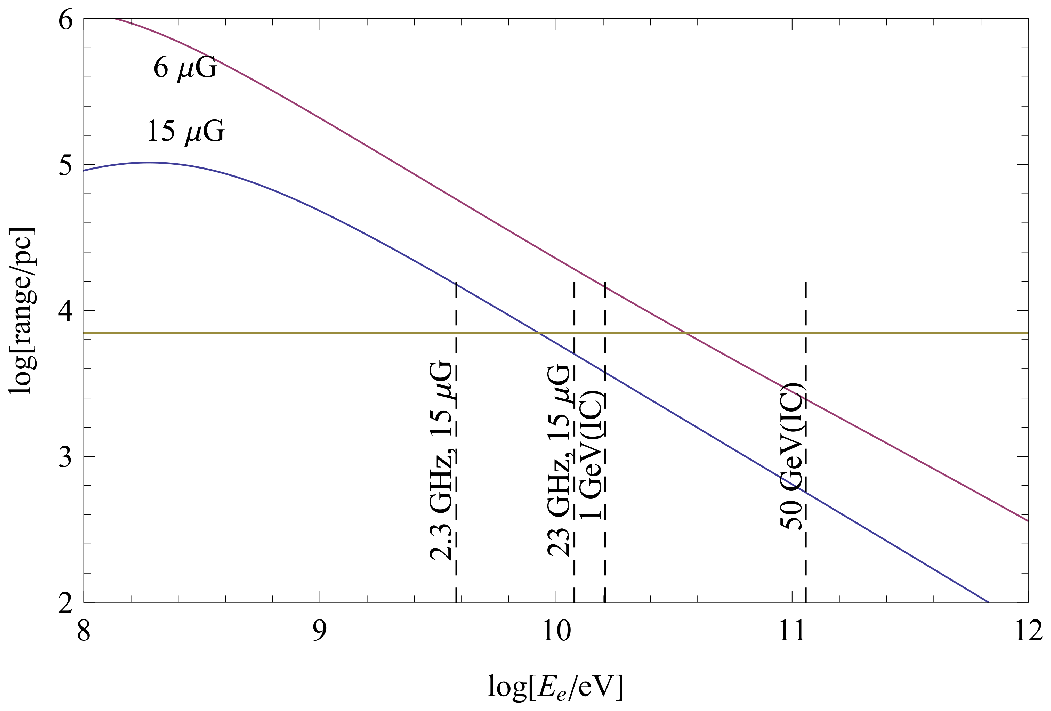}
\caption{
\label{fig_4}
}
\end{figure}

Three distinct emission ridges that all curve towards  
Galactic
west with increasing Galactic latitude are visible within the Lobes (Figure~\ref{fig_1}); 
two other substructures proceeding roughly north-west and south-west from around the Galactic Centre 
hint at limb brightening in the biconical base of the Lobes.
These substructures all have counterparts in WMAP polarization maps (Figure~\ref{wmap:Fig})
and one of them\cite{Jones2012}, already known from radio continuum data as the Galactic Centre Spur\cite{Sofue1989},   
appears to connect  back to the Galactic Centre; we label the other substructures  the Northern and Southern Ridges. 
The Ridges' magnetic field directions  (Figure~\ref{wmap:Fig}) curve following their structures.
The Galactic Centre Spur and Southern Ridges also seem to have GeV $\gamma$-ray counterparts (Figure~\ref{lobes:Fig} also cf. ref.[\cite{Su2012}]).
The two limb brightening spurs at the biconical Lobe base are also visible in the WMAP map, where they appear to connect back to the Galactic Centre area. A possible third spur develops north-east from the Galactic Centre. These limb brightening spurs are also obvious in the Stokes U map as an X-shape structure centred at the Galactic Centre (Figure~S3  of Supplementary Information).
Such coincident, non-thermal radio, microwave and $\gamma$-ray emission indicates the presence of a non-thermal electron population covering at least the energy range 1-100~GeV (Figure \ref{fig_4}) that is simultaneously synchrotron radiating at radio and microwave frequencies and up-scattering ambient radiation into $\gamma$-rays by the Inverse Compton process.
The widths of the Ridges are remarkably constant at $\sim$ 300 pc over their lengths.
The Ridges have  polarization fractions of 25 to 31\% (see Supplementary Information), 
similar to the average over the  Lobes. 
Given this emission and the stated polarization fractions, we infer magnetic field intensities   $6-12 \ \mu$G for the Lobes and $13-15 \ \mu$G  the Ridges
(see Figure~\ref{lobes:Fig} and ~\ref{wmap:Fig}  and the Supplementary Information).

An important  question about the Fermi Bubbles is whether they are ultimately powered by star-formation or by activity of the Galaxy's central, super-massive black hole.
Despite their very large extent, the $\gamma$-ray Bubbles and the  X-shaped 
 polarized microwave and X-ray 
structures tracing their limb-brightened base\cite{Bland-Hawthorn2003} have a narrow waist of only 100-200 pc diameter at the Galactic Centre. 
This matches the extent of the star-forming molecular gas ring ($\sim 3 \times 10^7 \msun$) recently demonstrated to occupy the region\cite{Molinari2011}.
With 5-10\% of the Galaxy's molecular gas content\cite{Morris1996}, star-formation activity in this `Central Molecular Zone'  is intense, accelerating a distinct cosmic ray population\cite{Aharonian2006,Crocker2011b} and driving an outflow\cite{Bland-Hawthorn2003,Law2010}  
of hot, thermal plasma, cosmic rays, and `frozen-in' magnetic field lines\cite{Crocker2010b,Crocker2011b,Crocker2011c}.

One consequence of the region's outflow is that the cosmic ray electrons accelerated there (dominantly energised by supernovae) are advected away before they lose much energy radiatively in situ\cite{Crocker2010a,Crocker2010b,Crocker2011b}.
This is revealed by the fact that the radio continuum flux on scales up to 800 pc around the Galactic Centre is in anomalous deficit with respect to the
 expectation afforded by the
 empirical far-infrared-radio continuum correlation\cite{Yun2001}.
The total 2.3~GHz radio continuum flux from the Lobes of $\sim$ 21~kJy, however, saturates this correlation normalising to the
inner $\sim 160$ pc  diameter region's 60 $\mu$m flux of 2 MJy[\cite{Launhardt2002}].
Together with the morphological evidence, this strongly indicates the Lobes are 
illuminated by cosmic ray electrons accelerated in association with star-formation within this region (see the Supplementary Information) 
not as a result of black hole activity.

The  Ridges appear to be continuous windings of individual, collimated
 structures around 
a general biconical outflow out of the Galactic Centre. 
The sense of Galactic rotation 
(clockwise as seen from Galactic north) and angular-momentum conservation  
mean that the Ridges get 
`wound-up'\cite{Heesen2011}  in the outflow with increasing distance from the plane
explaining the projected curvature of the 
visible, front-side of the
Ridges towards Galactic west. 
Polarized, rear-side emission is attenuated rendering it difficult to detect against the Lobes' front-side and the Galactic plane's stronger emission (Figure~\ref{fig_1} and the Supplementary Information).

For cosmic ray electrons synchrotron-emitting at 2.3~GHz  to be able to ascend to the top of the Northern Ridge at $\sim$ 7~kpc in the time it takes them to cool (mostly via synchrotron emission itself) requires vertical transport speeds of  $>$500 km/s (for  15 $\mu$G; see Figure~\ref{fig_4}).
Given the geometry of the GC Spur, the outflowing plasma is moving at $1,000-1,100$ km/s 
(Figure~\ref{fig_4} and the Supplementary Information), somewhat faster than the $\sim 900$ km/s gravitational escape velocity from the Galactic Centre region\cite{Muno2004},
implying that 2.3-GHz-radiating electrons can, indeed, be advected to the top of the Ridges in their loss time.

Given the calculated  fields and the speed of the outflow, the total magnetic energy  for each of the Ridges, $(4-9) \times 10^{52}$ erg  (see the Supplementary Information), is injected at a rate $\sim 10^{39}$ erg/s  over a few $ 10^6$ years; this is very close to the rate at which  independent modelling\cite{Crocker2011c} suggests  Galactic Centre star formation  is injecting magnetic energy into the region's outflow.
On the basis of the Ridges' individual energetics, geometry, outflow velocity,  timescales, and plasma content (see Supplementary Information), 
we suggest that their footpoints are energised by and rotate with the 
super-stellar clusters inhabiting\cite{Morris1996} the inner $\sim$100 pc (in radius) of the Galaxy.
In fact, we suggest that 
the Ridges constitute `phonographic' recordings of the last $\sim$10~Myr of Galactic Centre star-formation. 
Given its morphology, the Galactic Centre Spur likely still has an
active footprint. 
In contrast, the Northern and Southern Ridges seem not
to connect to the plane at 2.3 GHz. 
This may indicate their footpoints
are no longer active though the Southern Ridge may be connected to the
plane by a gamma-ray counterpart (see fig.~\ref{lobes:Fig}).
Unfortunately, present data do not allow us to trace the  Galactic Centre Spur all the way down to the plane but a connection between this structure and one (or some combination) of the 
$\sim1^\circ$-scale radio continuum spurs\cite{Pohl1992,Law2010} emanating north of the star-forming giant molecular cloud complexes Sagittarius B and C or the bright, non-thermal `Radio Arc'\cite{Morris1996} (itself longitudinally coincident with the $\sim$4 Myr Quintuplet\cite{Hussmann2011} stellar cluster), seems plausible.

The magnetic energy content of both Lobes is much larger than the Ridges, $(1-3) \times 10^{55}$ erg.
This suggests the magnetic fields of the Lobes are the result of the accumulation of a number of star formation episodes. 
Alternatively, if the Lobes' field structure were formed over the same timescale as the Ridges, it would have to be associated with recent activity of the super-massive black hole, perhaps occurring in concert with enhanced nuclear star-formation activity\cite{Zubovas2011}. 

Our data indicate
 the process of  gas accretion on to the Galactic nucleus inescapably involves  star-formation which, in turn, energises an outflow.
This carries away low angular momentum gas, cosmic rays and magnetic field lines and has a number of important consequences:
i)  The  dynamo activity in the Galactic Centre\cite{Ferriere2009}, likely required to generate its strong\cite{Crocker2010a} in situ field, requires the continual expulsion of small-scale helical fields to prevent dynamo saturation\cite{Brandenburg2005};
the presence of the Ridges high in the halo may attest to this process.
ii) The Lobes and Ridges reveal how the very active star formation in the Galactic Centre
 generates and sustains a strong, large-scale magnetic field structure in the Galactic halo. 
The effect of this on the propagation of high-energy cosmic rays  in the Galactic halo should be considered.
iii) The process of gas expulsion in the outflow
may explain how the Milky Way's super-massive black hole is kept relatively quiescent\cite{Morris1996} despite sustained, inward movement of gas.
%


\begin{addendum}
\item[Supplementary Information] is linked to the online version of the paper at www.nature.com/nature. 
 \item[Acknowledgements]
 This work has been carried out in the framework of the
S-band All Sky Survey collaboration (S-PASS).
We thank the Parkes Telescope staff for outstanding support both while setting up the non-standard observing mode and during the observing runs.
RMC thanks Felix Aharonian, Rainer Beck, Geoff Bicknell, David Jones, Casey Law,  Mark Morris,  Christoph Pfrommer, Wolfgang Reich, Andrea Stolte, Troy Porter, and Heinz V{\"o}lk for discussions and the Max-Planck-Institut f{\" u}r Kernphsik for supporting his research.
RMC also acknowledges the support of a Future Fellowship from the
Australian Research Council through grant FT110100108.
BMG and CP acknowledge the support of an Australian Laureate Fellowship from the
Australian Research Council through grant FL100100114.
MH acknowledges the support of the research programme
639.042.915, which is partly financed by the Netherlands Organisation for Scientifc Research (NWO). 
The Parkes radio
telescope is part of the Australia Telescope National Facility which is funded by the Commonwealth of Australia for
operation as a National Facility managed by CSIRO. 
\item[Author Contributions] EC performed the S-PASS observations, was the leader of the project, developed and performed the data reduction package,  did the main analysis and interpretation. 
RMC provided theoretical analysis and interpretation.
LSS,  MH, and SP performed the S-PASS observations.
MJK performed the telescope special setup that has allowed the survey execution.
LSS, MH, BMG, GB, MJK, and SP were co-proposers and contributed to the definition of the project.
CP performed the estimate of the H$_\alpha$ depolarizing region distance.
EC and RMC wrote the paper together. 
All the authors discussed the results and commented on the manuscript.  
 \item[Author Information] The authors declare that they have no
competing financial interests. Correspondence and requests for materials
should be addressed to E.C.~(email: 	Ettore.Carretti@csiro.au). 
Reprints and permissions information is available at npg.nature.com/reprintsandpermissions.
\end{addendum}

\pagebreak

\section*{Figure Captions}

{\bf Figure 1: Linearly polarized 
intensity $P$ at 2.3 GHz from S-PASS} ($P \equiv\sqrt{Q^2 + U^2}$).
   The thick dashed lines delineate the radio Lobes reported in this Letter,
   while the thin dashed lines delimit the 
   $\gamma$-ray Fermi Bubbles\cite{Su2010}. 
   The map is in Galactic coordinates, centred at the Galactic Centre with Galactic east to the left and Galactic north up; the Galactic Plane runs horizontally across the centre of the map.
      The polarized flux  density is indicated by the scale bar given in unit of Jy/beam with a beam size of 10.75' (1 Jy $\equiv 10^{-26}$ W m$^{-2}$Hz$^{-1}$).
   The Lobe edges follow  the $\gamma$-ray border up to $b \sim |30|^\circ$ where the radio emission extends beyond.
    The three polarized radio Ridges discussed in the text are also indicated along with the two limb brightening spurs.
The  Ridges appear to be the front side of a continuous windings of collimated
 structures around the general biconical outflow of the Lobes (see text). 
The Galactic Centre Spur  is nearly vertical at low latitude, possibly 
explained by a projection effect if it is mostly at
the front of the northern Lobe. 
At its higher latitudes, the Galactic Centre Spur becomes roughly parallel with the Northern Ridge (above) which itself exhibits little curvature; 
this is consistent with the overall outflow's becoming cylindrical above 4-5 kpc as previously suggested\cite{Bland-Hawthorn2003}.
In such a geometry,  synchrotron emission from the rear side of each cone is attenuated 
by a factor $\gsim 2$ with respect to the front-side, 
rendering it difficult to detect the former against the foreground of the latter and of the Galactic plane (see Supplementary Information).    

\pagebreak

{\bf Figure 2: Lobes' polarized intensity  and $\gamma$--ray spurs.} Schematic rendering of the edges of  two $\gamma$-ray substructures evident in the 2-5 GeV Fermi data as displayed in Figure~2 of ref.\cite{Su2010} which seem to be counterparts of the Galactic Centre Spur and the Southern Ridge.
The latter appears to be connected to the Galactic Centre by its $\gamma$-ray counterpart.
With the flux densities and polarization fraction quoted in the text we can infer  
equipartition\cite{Beck2005} magnetic field intensities of  $B_{eq} \sim 6 \ \mu$G  (1 $\mu$G $\equiv 10^{-10}$ T) if the synchrotron-emitting electrons occupy the entire volume of the Lobes, 
or $\sim 12 \ \mu$G if they occupy only a 300~pc thick skin (the width of the Ridges).
For the Southern Ridge, $B_{eq} \sim 13 \  \mu$G; for the Galactic Centre Spur $B_{eq} \sim 15 \ \mu$G; and, for the Northern Ridge,  $B_{eq} \sim 14 \ \mu$G.
   Note the large area of depolarization and small angular scale signal modulation visible across the Galactic plane extending up 
   to $b\sim |10|^\circ$ on either side of the Galactic Centre (thin dashed line). 
This depolarization is due to Faraday Rotation by a number of shells that match H$_\alpha$  emission regions\cite{Gaustad2001}, most of them lying in the Sagittarius arm at distances from the Sun up to 2.5~kpc, and some in the Scutum-Centaurus arm at $\sim$3.5~kpc. The small scale modulation is associated to weaker H$_\alpha$ emission encompassing the same HII regions and most likely associated with the same spiral arms. 
Thus 2.5~kpc  constitutes a lower limit to the Lobes' near side distance and places the far side beyond 5.5~kpc from the Sun (cf. ref.~\cite{Jones2012}).
Along with their direction in sky, this suggests the Lobes are associated with the Galactic Bulge and/or Centre. 

\pagebreak

{\bf Figure 3: Polarized intensity and magnetic angles at 23 GHz from WMAP[\cite{Hinshaw2009}]}. The magnetic angle is orthogonal to the emission polarization angle and traces the magnetic field direction projected on to the plane of the sky.
    The three ridges  are obvious while traces of the radio Lobes are visible (2.3 GHz edges shown by the black solid line). The magnetic field is aligned with the ridges and curves following their shape. Two spurs match the Lobe edges northwest and southwest of  Galactic Centre and could be limb brightening of the Lobes.  A third limb brightening spur candidate is also visible north east of the Galactic Centre.
The map is in Galactic coordinates, centred at the Galactic Centre. Grid lines are spaced by $15^\circ$.
 The emission intensity is in Brightness Temperature, the unit is K. Data have been binned in 1$^\circ \times 1^\circ$ pixels to improve the signal-to-noise ratio.
From a combined analysis of microwave 
and $\gamma$-ray data (see also the Supplementary Information) we can derive the following magnetic fields limits (complementary  to the equipartition limits reported in the text and Figure~2): 
for the overall Lobes/Bubbles $B > 9 \ \mu$G and for the Galactic Centre Spur  $11 \ \mu$G $< B < 18 \ \mu$G.  

\pagebreak

{\bf Figure 4: The vertical range of cosmic ray electrons as a function of their kinetic energy. }
(Due to geometrical uncertainties, adiabatic losses cannot be determined so that the plotted range actually constitutes an upper limit.)
Electrons are taken to be transported with a speed given by the sum of the inferred vertical wind speed (1100 km/s) and the vertical component of the Alfven velocity in the magnetic field.
The former is inferred from the geometry of the Northern Ridge:  if its footpoint has executed roughly half an orbit in the time the Galactic Centre Spur has ascended to its total height of $\sim 4$ kpc, its upward velocity 
must be close to 1,000~km/s$\times (r/100 \ \textrm{pc})^{-1} \times v_{rot}/(80 \ \textrm{km/s})$[\cite{Jones2012}], where we have normalised to a footpoint rotation speed of 80 km/s at a radius of 100 pc from the Galactic Centre\cite{Molinari2011} (detailed analysis gives 1100 km/s: see the Supplementary Information).
In a strong, regular magnetic field, the electrons are expected to stream ahead of the gas at the Alfven velocity\cite{Kulsrud1969} in either the Ridges
($B  \simeq 15 \ \mu$G, $v_A^{vert} \simeq 300$ km/s; this is a lower limit given 
that $n_H \lsim 0.008$ cm$^{-3}$ on the basis of the ROSAT data\cite{Almy2000}) or in 
the large-scale field of the Lobes ($B  \simeq 6 \ \mu$G, $v_A^{vert} \gsim 100$ km/s for $n_H \lsim 0.004$ cm$^{-3}$ in the Lobes' interior as again implied by the  data).
Also plotted are: the characteristic energies of electrons synchrotron radiating at 2.3 and 23 GHz (for a $15 \ \mu$G field) and into 1 and 50 GeV $\gamma$-rays via inverse Compton (`IC') upscattering of a photon background with characteristic photon energy 1 eV; and the approximate 7 kpc distance of the top of the Northern Ridge from the Galactic plane.


\pagebreak

\pagebreak

\pagebreak

\pagebreak

\end{document}